\documentclass[a4paper,12pt]{article}
\usepackage{amsmath}
\usepackage{citehack}
\usepackage{bm}
\usepackage{graphicx}
\usepackage{epstopdf}
\usepackage{color}
\usepackage{comment}
\usepackage{hyperref}
\usepackage{float}
\usepackage{tikz, xcolor, pgfplots}
\usetikzlibrary{shapes, arrows}

\begin{document}

\title{Molecular-dynamic simulation of water vapor interaction with suffering pores of the cylindrical type\thanks{topic JINR LIT No. 05-6-1118-2014/2019, protocol No. 4596-6-17/19.}}
\author{E.G. Nikonov$^1$, M.Pavlu\v{s}$^2$, M. Popovi\v{c}ov\'a$^2$\\
\begin{minipage}{10cm}
\begin{center}
\small\em
\vspace{4mm}
$^1$Joint Institute for Nuclear Research,\\ 141980 Dubna, Moscow Region, Russia\\ email: e.nikonov@jinr.ru \\
\vspace{4mm}
$^2$University of Pre\v{s}ov,\\ str. Kon\v{s}tantinova 16, 080 01 Pre\v{s}ov,  Slovakia\\ email: miron.pavlus@unipo.sk, maria.popovicova@unipo.sk
\end{center}
\end{minipage}
} 
\date{}
\maketitle 

\begin{abstract}
Theoretical and experimental investigations of water vapor interaction with porous materials are carried out both at the macro level and at the micro level. At the macro level, the influence of the arrangement structure of individual pores on the processes of water vapor interaction with porous material as a continuous medium is studied. At the micro level, it is very interesting to investigate the dependence of the characteristics of the water vapor interaction with porous media on the geometry and dimensions of the individual pore.

In this paper, a study was carried out by means of mathematical modelling of the processes of water vapor interaction  with suffering pore of the cylindrical type. The calculations were performed using a model of a hybrid type combining a molecular-dynamic and a macro-diffusion approach for describing water vapor interaction with an individual pore. The processes of evolution to the state of thermodynamic equilibrium of macroscopic characteristics of the system such as temperature, density, and pressure, depending on external conditions with respect to pore, were explored. The dependence of the evolution parameters on the distribution of the diffusion coefficient in the pore, obtained as a result of molecular dynamics modelling, is examined. The relevance of these studies is due to the fact that all methods and programs used for the modelling of the moisture and heat conductivity are based on the use of transport equations in a porous material as a continuous medium with known values of the transport coefficients, which are usually obtained experimentally.
\end{abstract}

\section{Introduction}
One of the most important problem for theoretical and experimental investigations of water vapor interaction with porous materials is how to combine calculation results obtained by micro level study, for example molecular dynamic and Monte-Carlo simulation, analytical investigations by solving mass transfer equations   for single pore etc. \cite{Alim2017} with  experimental data and calculations by empirical formulas which is usually presented by macro characteristics such as porosity, density etc. \cite{krus}. 
At the macro level, the influence of the arrangement structure of individual pores on the processes of water vapor interaction with porous material as continuous medium is ordinary studied. 
But even at the level of individual pore molecular dynamic simulation data does not directly correspond to macro characteristics of water vapor interaction with a pore. So if macro equations like, for example, diffusion equations can be solved for some geometry of the pore it would be useful to use the following two steps solution scheme. At the first step we can obtain some characteristics like, for example, diffusion coefficient by molecular dynamic simulations. At the second step we can use the values of these diffusion coefficients for solving diffusion equations at the macro level. Moreover, the dependence of the characteristics of the water vapor interaction with porous media on the geometry and dimensions of the individual pore can be investigated at the micro level, in more simple way then in the case of diffusion or mass transfer equations, at macro level.

In this paper, processes of water vapor interaction  with suffering pore of the cylindrical type are considered. Investigations are carried out by means of mathematical modelling. The calculations are performed using hybrid model combining molecular dynamic and a macro-diffusion approach for describing water vapor interaction with individual pore, like in \cite{NPP1709} and \cite{NPP1708}. The processes of evolution to the state of thermodynamic equilibrium of macroscopic characteristics of the system such as temperature, density, and pressure, depending on external conditions with respect to pore, is explored. The dependence of the evolution parameters on the distribution of the diffusion coefficient in the pore, obtained as a result of molecular dynamics modelling, is researched. The topicality of these explorations is due to the fact that all methods and programs used for the modelling of the heat and moisture conductivity are based on the use of transport equations in porous material as continuous medium with known values of transport coefficients. These coefficients are usually obtained experimentally or using empirical formulas. Analysis of the following modern software for Modelling of Heat, Air, Moisture (HAM) transfer through porous media is carried out. 
\begin{itemize}
\item The Heat, Air and Moisture Tool Kit (Quirouette Building Science Software, http://www.qbstoolbox.com/).
\item NRC-IRC hygIRC-1-D: helps design community choose optimal building envelope components and systems (https://www.nrc-cnrc.gc.ca/eng /projects/irc/hygirc.html).
\item  WUFI - Oak Ridge National Laboratory (ORNL)/Fraunhofer IBP is a menu-driven PC program which allows realistic calculation of the transient coupled one-dimensional heat and moisture transport in multi-layer building components exposed to natural weather. It is based on the newest findings regarding vapor diffusion and liquid transport in building materials and has been validated by detailed comparison with measurements obtained in the laboratory and on outdoor testing fields.
\item COMSOL (https://www.comsol.com) for Modeling Heat and Moisture Transport in Building Materials is  also very powerful and multifunctional software. 
\end{itemize}
All of these programs use calculation methods for modelling of moisture transfer through porous media based on transport differential equations with previously known transport coefficients. So it is necessary to know values of the transport coefficients before using of these programs. One of the alternative ways  with respect to the experience is to obtain these coefficients by molecular dynamic simulations. We present this alternative 
in this paper.
\section{Molecular dynamics model}
\label{mdm}
In classical molecular dynamics, the behavior of an individual particle is described by the Newton equations of motion \cite{Gould}, which can be written in the following form
\begin{equation}\label{a} 
  m_i\frac{d^2 \vec{r_i}}{dt^2}=\vec{f_i},
\end{equation}
\noindent
where $i \ - $ a particle number, $(1\leq i \leq N)$, $N \ - $ the total number of particles, $m_i \ - $ particle mass, $\vec{r_i}\ - $ coordinates of position,  $\vec{f_i} \ - $ the resultant of all forces acting on the particle. This resultant force has the following representation
\begin{equation}\label{b} 
  \vec{f_i} = -\frac{\partial U(\vec{r_1},\ldots,\vec{r_N})}{\partial \vec{r_i}} + \vec{f_i}^{ex},
\end{equation}
where $U \ - $ the potential of particle interaction, $\vec{f_i}^{ex} - $ a force caused by external fields. 
For a simulation of particle interaction, we use the Lennard-Jones potential \cite{LJ} with $\sigma = 3.17 \mbox{\AA}$ and $\varepsilon = 6.74\cdot 10^{-3}$ eV. It is the most used to describe the evolution of water in liquid and saturated vapor form. Equations of motion 
(\ref{a}) 
were integrated by Velocity Verlet method \cite{Verlet1967}.
Berendsen thermostat \cite{Berendsen1984} is used for temperature calibration and control. The coefficient of the velocity recalculation $\lambda(t)$ at every time step $t$ depends on the so called ''rise time'' of the thermostat $\tau_B$ which belongs to the interval $ [0.1,2] \ \mbox{psec}$. 
$\tau_B$ describes strength of the coupling of the system to a hypothetical heat bath. For increasing $\tau_B$, the coupling weakens, i.e. it takes longer to achieve given temperature $T_0$ from current temperature $T(t).$
The Berendsen algorithm is simple to implement and it is very efficient for reaching the desired temperature from far-from-equilibrium configurations. One of the main drawback of the Berendsen algorithm is that the ensemble generated when using the Berendsen thermostat is not a canonical ensemble\cite{Hberg} . The Berendsen thermostat is extremely efficient for relaxing a system to the target temperature, but once your system has reached equilibrium, it might be more important to probe a correct canonical ensemble.
Another method was originally introduced by Nose\cite{Nose1984} and subsequently developed by Hoover\cite{Hoover1985}. The idea is to consider the heat bath as an integral part of the system by addition of an artificial variable $ \tilde{s} $, associated with a ''mass'' $Q > 0$ as well as a velocity $ \dot{\tilde{s}} $. The magnitude of $Q$ determines the coupling between the reservoir and the real system an so influences the temperature
fluctuations. The artificial variable $ \tilde{s} $ plays the role of a time-scaling parameter, more precisely, the timescale in the extended system is stretched by the factor $ \tilde{s} $
$$
d\tilde{t} =  \tilde{s}dt.
$$

According to the paper\cite{Hberg} it is used the following thermostatization scheme for molecular dynamic simulations. Berendsen thermostat with a sufficiently small value of $\tau$ is applied for equilibration of water vapor directly in the pore. The Nose-Hoover thermostat with finite but not too large value of ''mass'' $Q$ is used for outer space of modelling system. This thermostatization scheme allows to control system temperature for canonical ensemble.
\section{Molecular dynamic model calibration}\label{mdmc}
\begin{figure}[H]
\begin{center}
\includegraphics[angle=-90,width=0.48\linewidth]{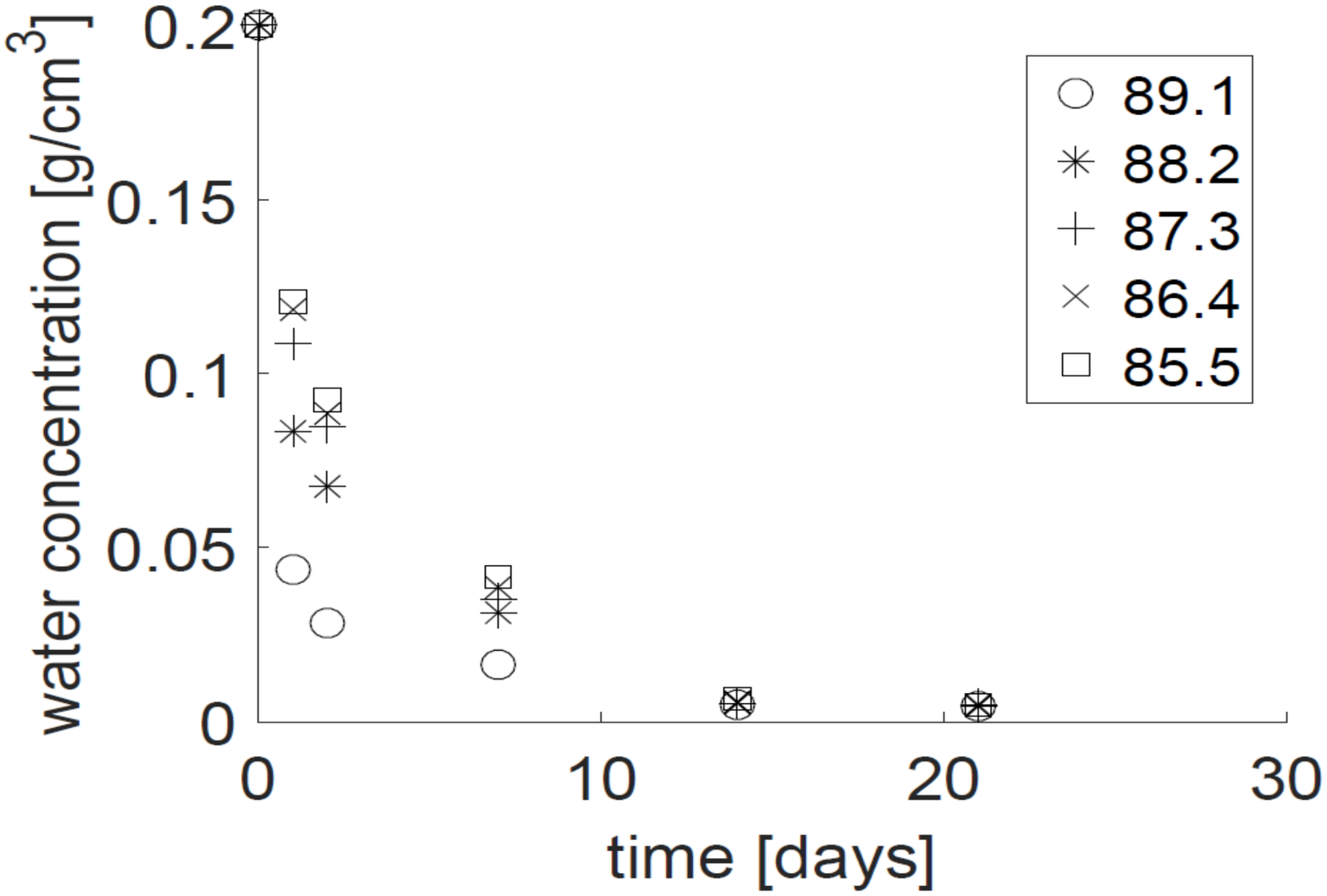}~
\includegraphics[angle=-90,width=0.48\linewidth]{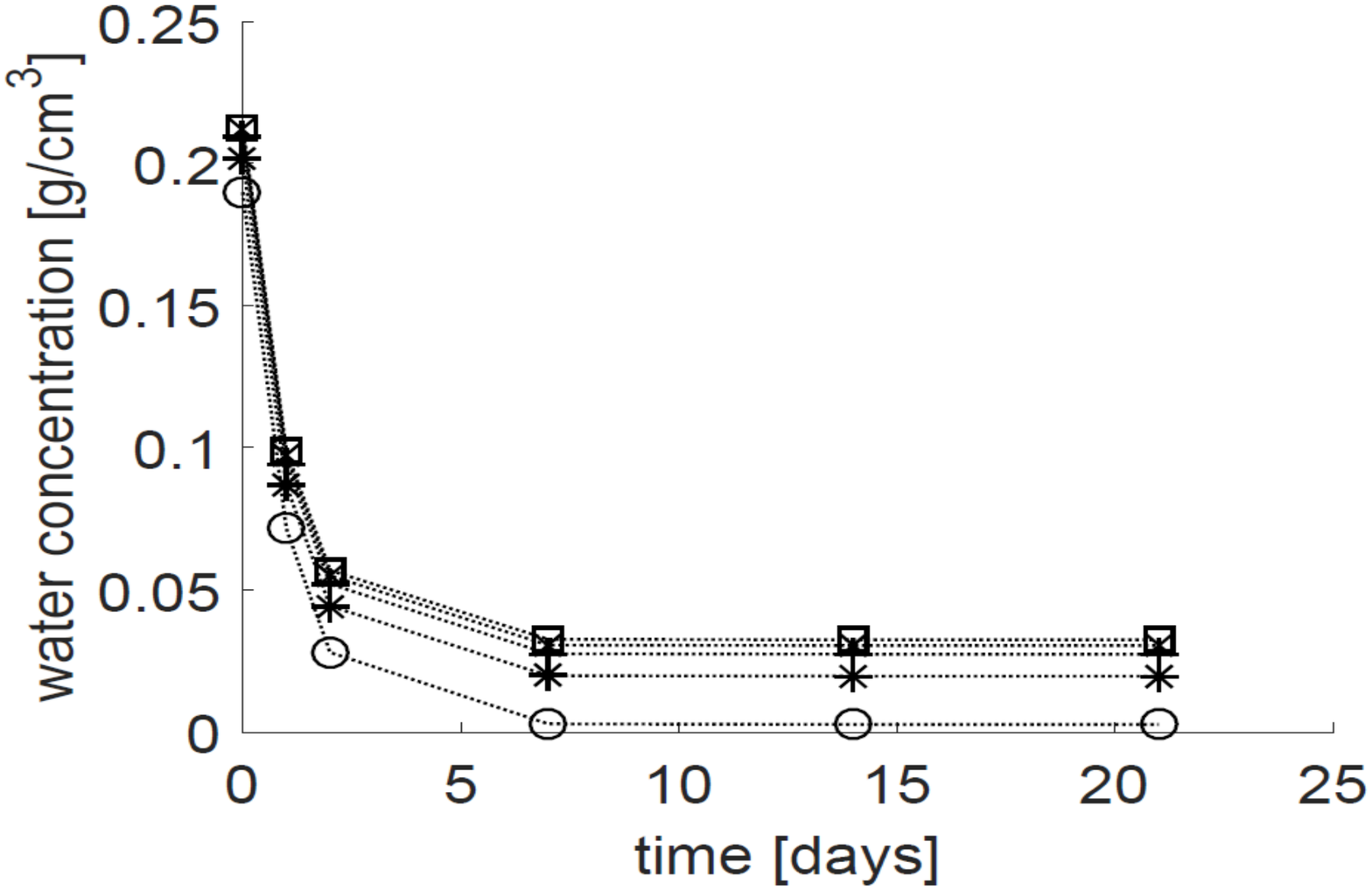}
\end{center}
\caption{Dynamics of water concentration at boundary.}
\label{hum_pic}
\end{figure}

\begin{figure}[H]
\begin{center}
\includegraphics[angle=-90,width=0.48\linewidth]{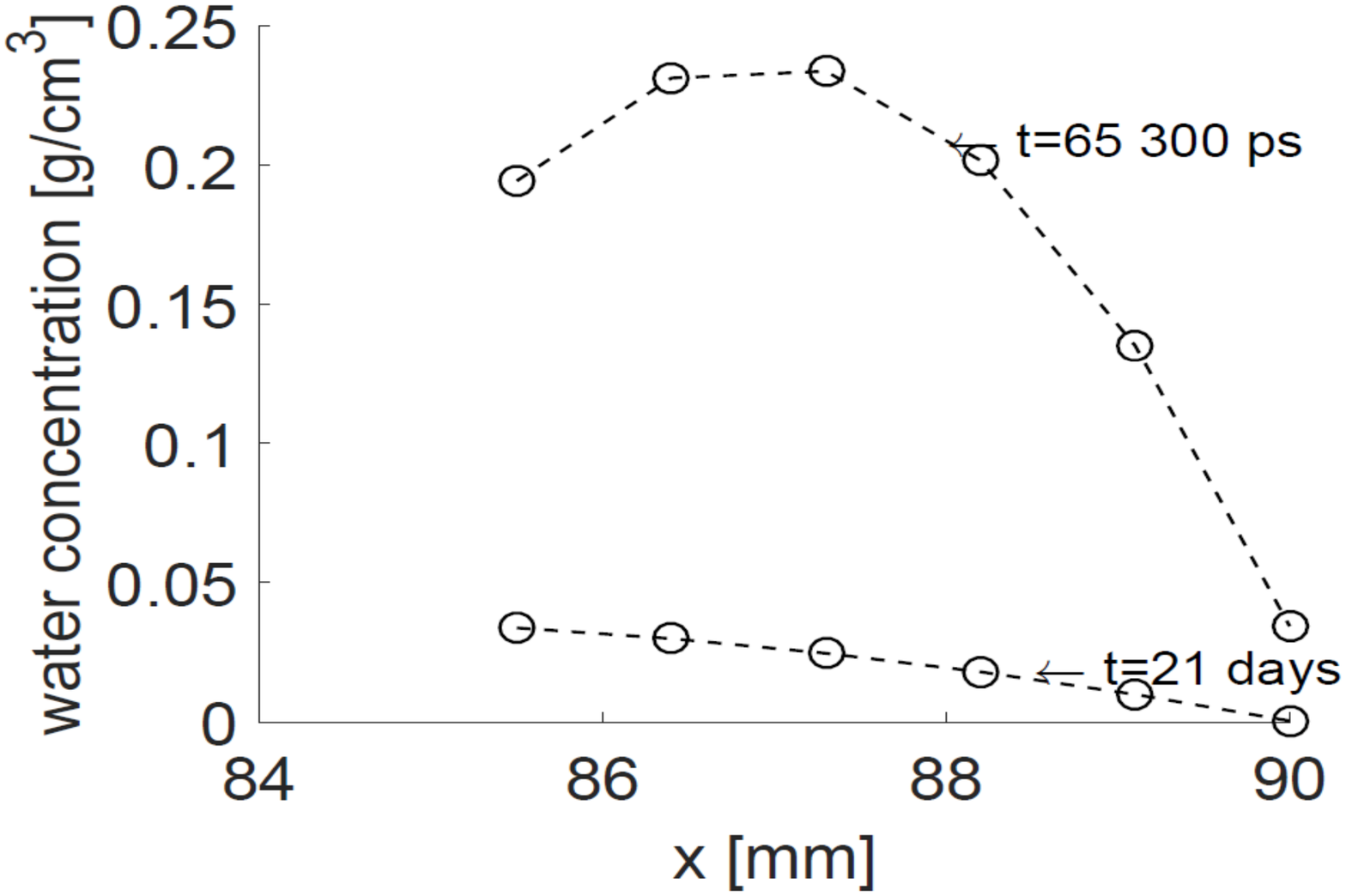}
\end{center}
\caption{Water concentration on the x.}
\label{hum_pic_x}
\end{figure}

We used experimental data for the investigation of the time and space dependent moisture distribution in a drying brick sample\cite{Pleinert1998} to adjust the values of the parameters $\tau_B,$  $Q$ and $\tilde{s}$ in our thermostatization scheme. The sample dimensions are $12\times 9\times 3$~cm$^3.$ All faces of the brick except for one face are isolated from the surrounding space by means of aluminum foil. Therefore evaporation can occur only through one open face along the length $9\,cm$. The open face is  $12\times 3$~cm$^2.$ Similarly to work\cite{Amirkhanov2008}, we take values of water concentration in points $x=89.1; 88.2; 87.3; 86.4; 85.5$~mm from the side of the open edge of the brick using the humidity distribution plot\cite{Pleinert1998}. We get the following plot for these points at the left picture and corresponding approximation by the function $ w = a + be^{-t}$ at the right picture (Fig. \ref{hum_pic}).
Consider the distribution of moisture at the boundary of the open edge at the point $x = 90\ $ mm - 250\ nm $=$ $ 89,999750\ $ mm, for two times $t=65300$  psec and $t=21$ days. Here, $t=21$ days is an experimental time and $t=t_0,$   $t_0=65300$ psec is our simulation time. Approximation by the parabola $w=ax^2+bx+c$ leads to the following distribution of water concentration for these two time moments (Fig. \ref{hum_pic_x}). Note, that $250$~nm presents a middle of a pore
that we will later consider. 
Thus, we get the following values for water concentration on the $x=89,999750$ mm layer. For $t=21$ days - $w = 0.00041511$~g/cm$^3$ and for $t=t_0$ - $w = 0.034386$~g/cm$^3.$ The last value for water concentration, we use for calibration of $\tau_B,$  $Q$ and $\tilde{s}$ parameters for corresponding thermostats. We use the ratio of water concentration at time $t=0$ and $t=t_0$ i.e. $K=0.2/0.034386=5.816321$ to calculate the number of water vapor molecules in the pore.   

\section{Macroscopic diffusion model}\label{Macro}
Let us denote the water vapor concentration as $w_v(r,\varphi,z,t)$ [$ ng/(nm)^3$] in the point $(r,\varphi,z,t)$ where $r,\varphi,z$ are space independent cylindrical variables and $t$ is time independent variable. Then, we consider the following macroscopic diffusion model
\begin{equation}\label{eq01}
\frac{\partial w_v}{\partial t}=D\Big[\frac{1}{r}\frac{\partial}{\partial r}\Big(r\frac{\partial w_v}{\partial r}\Big)+
\frac{1}{r^2}\frac{\partial^2 w_v}{\partial\varphi^2}+\frac{\partial^2 w_v}{\partial z^2}\Big]
\end{equation}
$$
\qquad
0<r<r_0\qquad 0<\varphi<2\pi\qquad 0<z<z_0\qquad  t>0
$$ 
\begin{equation}\label{eq02}
w_v(r,\varphi,z,0)=w_{v,0}\qquad 0\leq r\leq r_0\qquad 0\leq \varphi< 2\pi\qquad 0\leq z\leq z_0 
\end{equation} 
\begin{equation}\label{eq03}
\frac{\partial w_v}{\partial r}(r_0,\varphi,z,t)=0\qquad 0\leq \varphi< 2\pi\qquad 0\leq z< z_0\qquad t>0
\end{equation}
\begin{equation}\label{eq04}
\frac{\partial w_v}{\partial z}(r,\varphi,0,t)=0\qquad 0\leq r< r_0\qquad 0\leq \varphi< 2\pi\qquad t>0
\end{equation}
\begin{equation}\label{eq05}
-D\frac{\partial w_v}{\partial z}(r,\varphi,z_0,t)=\beta [w_v(r,\varphi,z_0,t)-w_{v,out}(t)]
\end{equation}
$$
\qquad 0\leq r< r_0\qquad 0\leq \varphi< 2\pi\qquad t>0.
$$ 
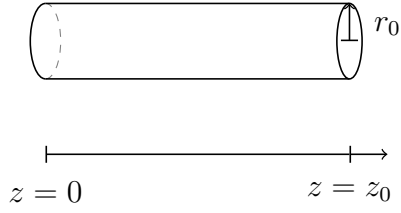
\begin{figure}[h]
\begin{center}
\begin{tikzpicture} 
\draw[dashed,color=gray] (0,1) arc (-90:90:0.2 and 0.5);
\draw[semithick] (0,1) -- (4,1);
\draw[semithick] (0,2) -- (4,2);
\draw[semithick] (0,1) arc (270:90:0.2 and 0.5);
\draw[semithick] (4,1.5) ellipse (0.166 and 0.5);
\draw[|->,semithick] (4,1.5) -- (4,2);
\draw (4.5,1.7) node {$r_0$};
\draw[|-,semithick] (0,0) -- (4,0);
\draw[|->,semithick] (4,0) -- (4.5,0);
\draw (0,-0.5) node {$z=0$};
\draw (4,-0.5) node {$z=z_0$};
\end{tikzpicture}
\end{center}
\caption{Shape of 3D pore.} \label{fig1}
\end{figure}
\noindent
where $D$ is the diffusion coefficient [$(nm)^2/psec$]; $r_0, 2\pi, z_0$ are 3D pore cylindrical dimensions [$nm$]; $w_{v,0}$ is the initial concentration
of water vapor; $\beta$ is the coefficient of water vapor transfer from pore space to outer space [$nm/psec$]; $w_{v,out}(t)$ is the water vapor concentration in outer space [$ng/(nm)^3$].

Because the initial (\ref{eq02}) and boundary (\ref{eq03})--(\ref{eq05}) conditions do not depend on the variable $\varphi,$ the solution of the problem (\ref{eq01})--(\ref{eq05}) does not depend on $\varphi,$ so that $w_v=w_v(r,z,t).$

We suppose that the outer space water vapor concentration is expressed as
$$
w_{v,out}(t)=\varphi_0\cdot w_{sv}(T_0), 
$$       
where $\varphi_0$ is the relative humidity of outer space ($0\le\varphi_0\le 1$) and $w_{sv}(T_0)$ is saturated water vapor concentration at outer temperature $T_0.$ 

In this case, the linear problem (\ref{eq01})--(\ref{eq05}) can be solved exactly by means of the variables separation method 
\cite{Bitsadze1980} and the result of the solution is the following 
\begin{multline}\label{eq06}
w_v(r,z,t)=w_{sv}(T_0)\cdot\varphi_0+\Big[w_{v,0}-w_{sv}(T_0)\cdot\varphi_0\Big]\cdot \\ \sum_{m=1}^{\infty}\sum_{n=0}^{\infty}e^{D\lambda_{mn} t}c_{mn}J_0(\alpha_{rn}r)\cos(\alpha_{zm} z)
\end{multline}
$$
\qquad 0\leq r\leq r_0\qquad 0\leq z\leq z_0\qquad t\ge0.
$$
Here, $c_{mn}$ are coefficients of unity expansion
$$
c_{mn}=
  \begin{cases}
    \frac{4\sin(\alpha_{zm} z_0)}{2 z_0\alpha_{zm}+\sin(2\alpha_{zm} z_0)}       & \quad \text{if } n=0;\ \ m=1,2,3,\dots\\
    0                                                                                                                          & \quad \text{if } n=1,2,3,\dots;\ \ m=1,2,3,\dots\\
  \end{cases}
$$
and $\lambda_{mn}$ are eigenvalues where 
$$
\lambda_{mn}=-\alpha_{zm}^2-\alpha_{rn}^2  
$$
and $\alpha_{zm},$ $\alpha_{rn}$ are solutions of the equations
$$
\alpha_{zm}\cdot\tan(\alpha_{zm} l_x)=\beta/D,\qquad m=1,2,3,\dots 
$$
$$
J'_0(\alpha_{rn}r_0)=0\qquad n=0,1,2,\dots.
$$
with $J_0(x)$ as the Bessel function of the first kind of zero order. 
\section{Computer simulation of micro model}\label{microsim}
Our micro model is made up of a pores in the shape of a cylinder with the radius $r_0=28.5$ nm and the length $z_0=500$ nm, the outer environment is simulated by a prism that is also $500$ nm long and its sides are $150$ nm, that is, the outer volume is 9 times greater than the cylinder volume.

Initial concentrations were obtained from the density of water vapor at the appropriate pressure and density at a given temperature using known tabulated data. The pressure in the pore was controlled using the formula based on virial equation \cite{Frenkel2002}.
$$
P = \frac{1}{3V}\left(\left\langle 2K\right\rangle-\left\langle\sum\limits_{i<j} r_{ij}\cdot f\left(r_{ij}\right)\right\rangle\right).
$$
Here $V$ is the pore volume, $\left\langle 2K\right\rangle$ is the doubled kinetic energy averaged over the ensemble, $f\left(r_{ij}\right)$ is the force between  particles $i$ and $j$ at a distance $r_{ij}$. 

At a temperature $T_0=25$ $^o$C and pressure $p_0=3.17$ kPa, 1000 molecules of water vapor are placed in the pore and 1800 molecules in the outer environment. The molecules are evenly distributed both in the pore and in the external environment. This means, we have 100 $\%$ saturated water vapor at a pressure $p_0$ in the pore and 20 $\%$ of saturated water vapor in the outside. At the beginning moment $t=0,$ we give the particles random velocities from the interval (-1, 1) and set the total velocity equal to zero.  Then, the velocities are rescaled so that the desired initial kinetic energy is achieved for both the pore and the outside.

Under these conditions at time $t=t_0,$ we expected $N=338$ molecules of water vapor in the pore resulting from the equation $\frac{1000-200}{N-200}=K$ where 200 molecules represent an outer concentration in our model.  



To achieve the desired value at $t=t_0,$ we use the above-mentioned thermostatization scheme for molecular dynamic simulations with the following values $\tau_B=0.66 \,ps$  for the Berendsen thermostat and $Q=100$ and $\tilde{s}(0)=0.1$ for the Nose-Hoover thermostat. As a result at the time $t=t_0,$ we have $341$ particles in our model.

\begin{figure}[ht]
\center{\includegraphics[angle=-90, width=0.48\linewidth]{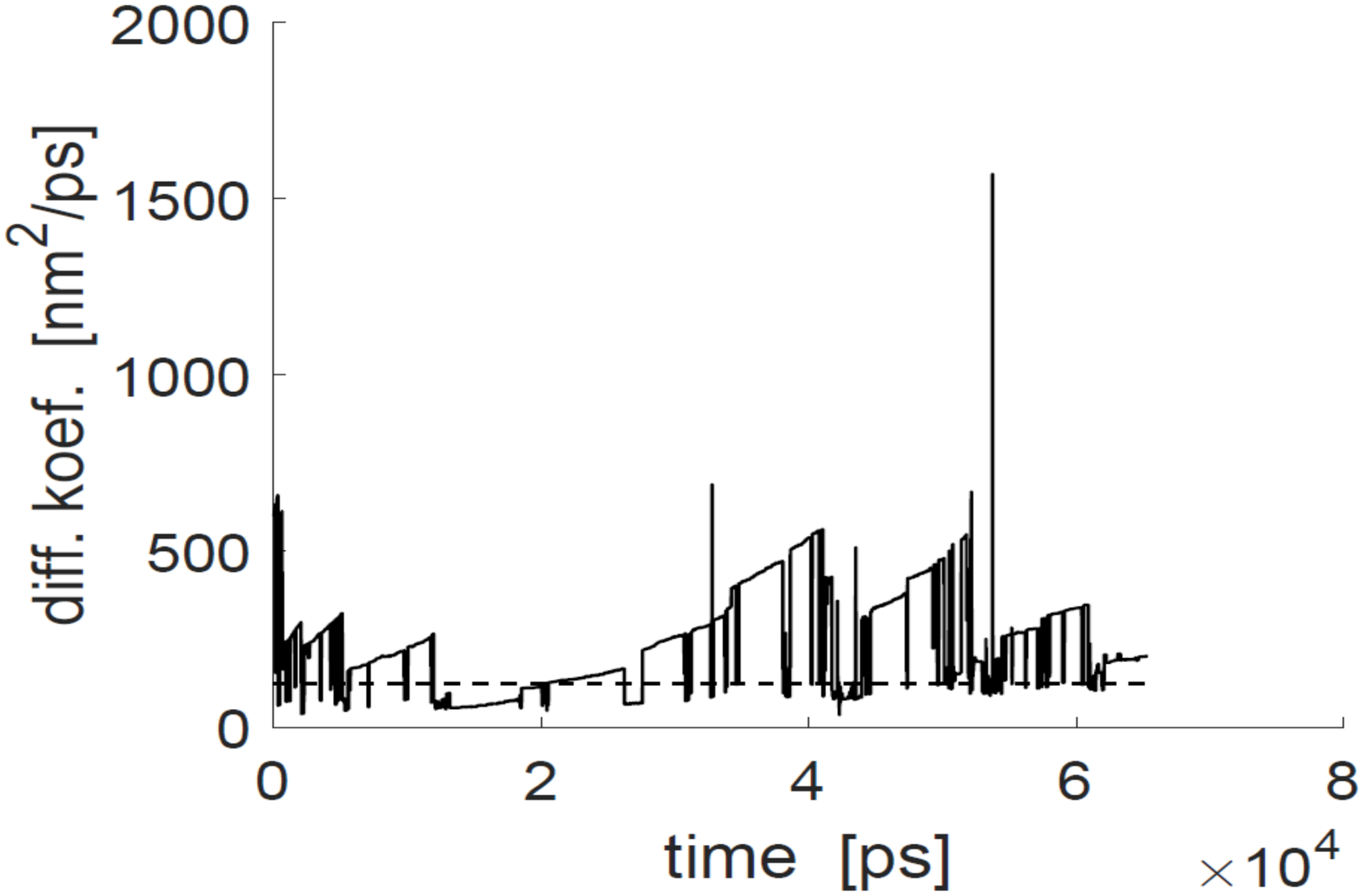}
\includegraphics[angle=-90, width=0.48\linewidth]{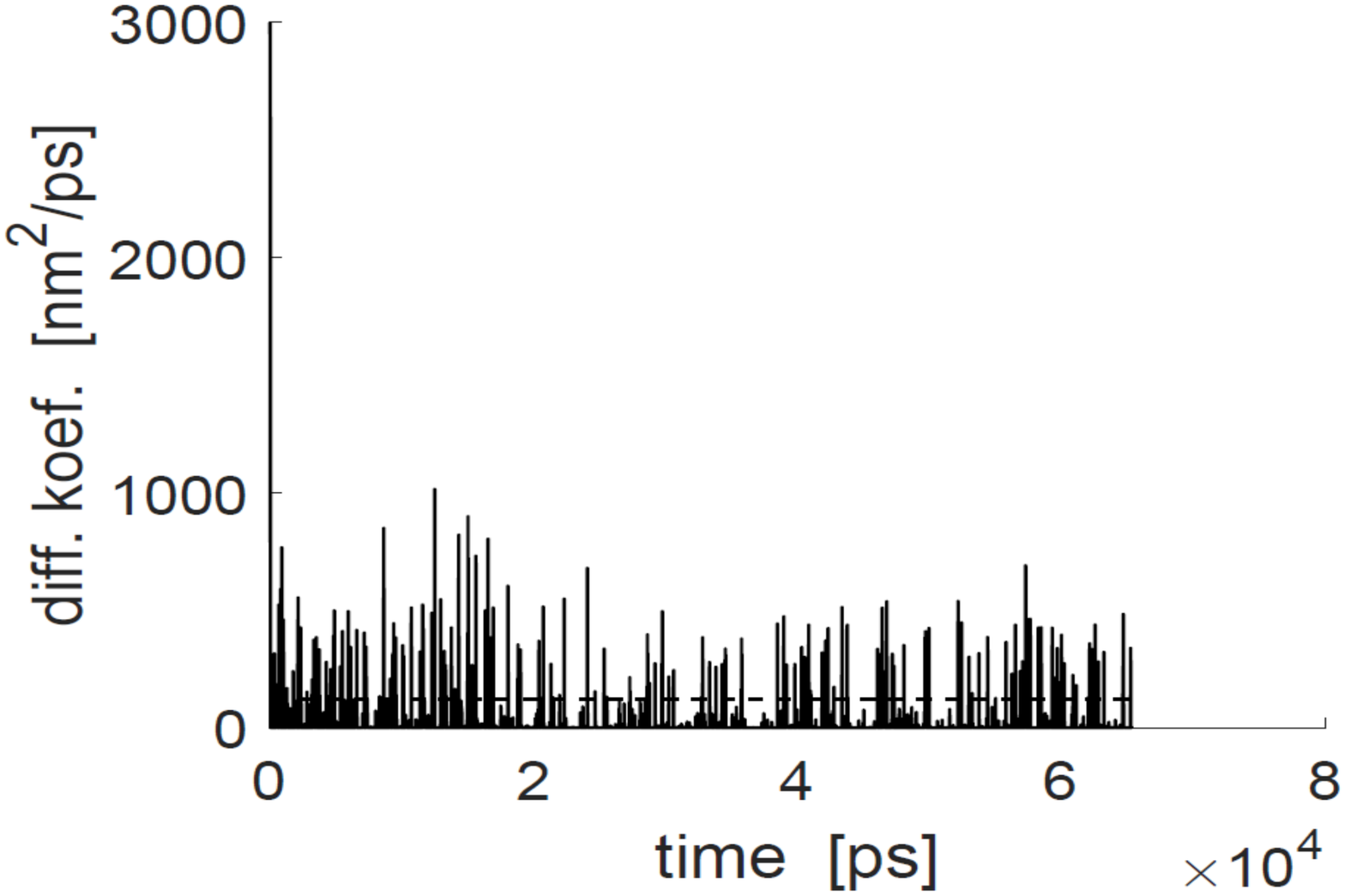}}
\caption{Zigzag diffusion coefficients for pore (left) and for outer space (right) of the drying process, and  constant value $D=122.39$ [nm$^2$/psec] (horizontal dashed line).}\label{Diffusion_Coef}
\end{figure}

The main characteristic of the diffusion process is the diffusion coefficient. One goal of the micro model is to construct a constant diffusion coefficient $D$ which will then be used in the macro-model. The constant value $D$ is a mean of the diffusion coefficient of the pore $D_{in}(t)$ and the diffusion coefficient of the outer space $D_{out}(t)$ that is calculated according to the formula (\ref{eq07})
\begin{equation}\label{eq07}
D=\frac{1}{t_0}\int_0^{t_0}\frac{D_{in}(t)+D_{out}(t)}{2}dt. 
\end{equation}
In the Fig. \ref{Diffusion_Coef} (left), we show the diffusion coefficients for both the pore $D_{in}(t)$ (zigzag curve) and the constant value $D=122.39$ [nm$^2$/psec] (horizontal dashed line). In the Fig. \ref{Diffusion_Coef} (right), we show the diffusion coefficients for both the outer space $D_{out}(t)$ (zigzag curve) and the same constant value $D=122.39$ [nm$^2$/psec] (horizontal dashed line). 

\begin{figure}[ht]
\center{
\includegraphics[angle=-90, width=0.48\linewidth]{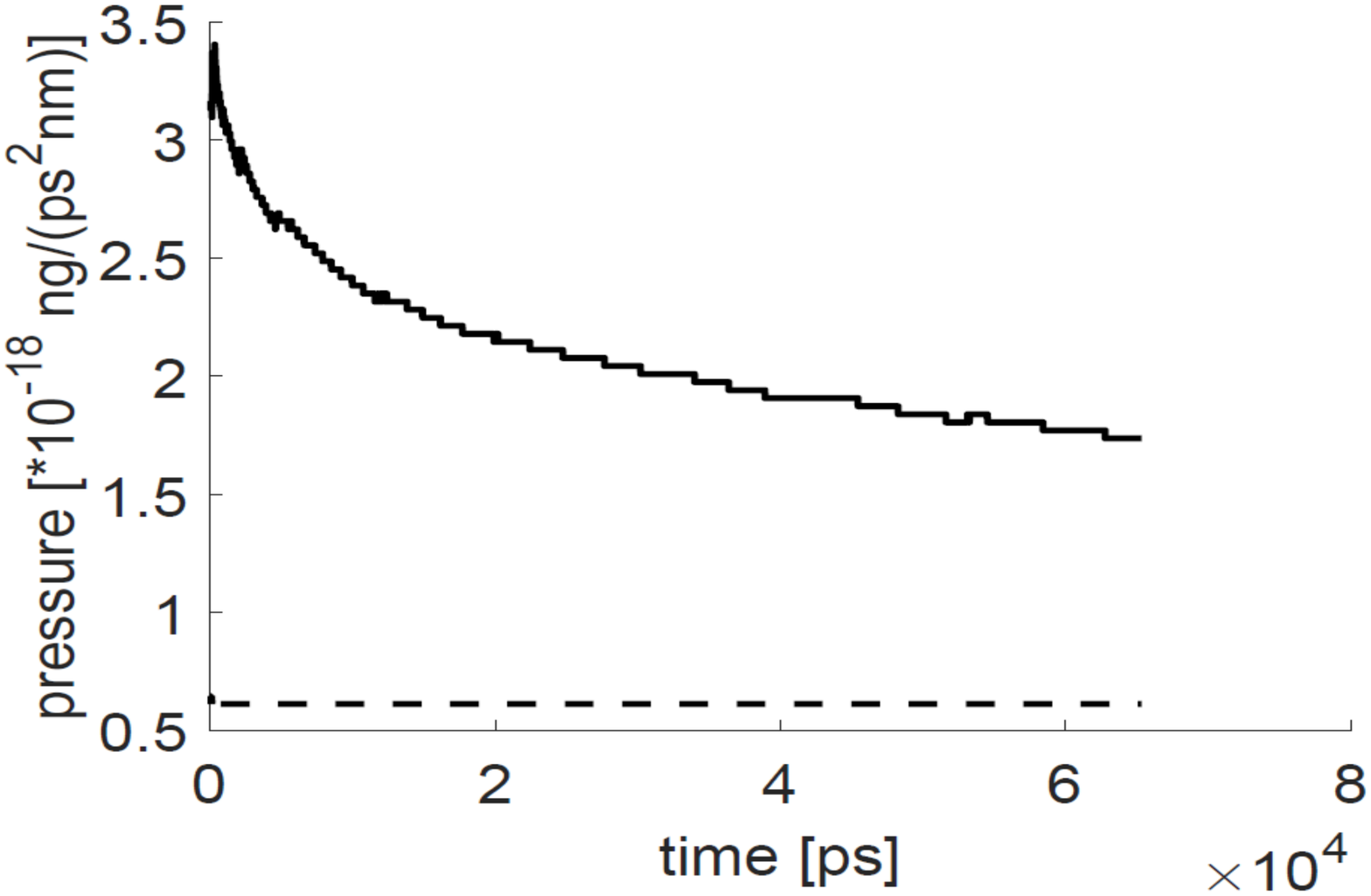}
}
\caption{Pressures  in the pore (up curve) and in the outer space (down curve) of the drying process.}\label{Pressures}
\end{figure}

Next in the Fig. \ref{Pressures}, we present dynamics of the pressures in the pore and in the outer space.   

\section{Computer simulation of macro-model}\label{macrosim}
In previous part, we showed the results of the micro model for the case of the saturated water vapor at temperature $T_0=25$ $^oC$ and at pressure $p_0=3.17$ $kPa$ and the outer space that is 9 times greater than the cylinder pore. We also received the constant diffusion coefficient $D=122.39$ [nm$^2$/psec].  

\begin{figure}[H]
\begin{center}
\includegraphics[angle=-90,width=0.48\linewidth]{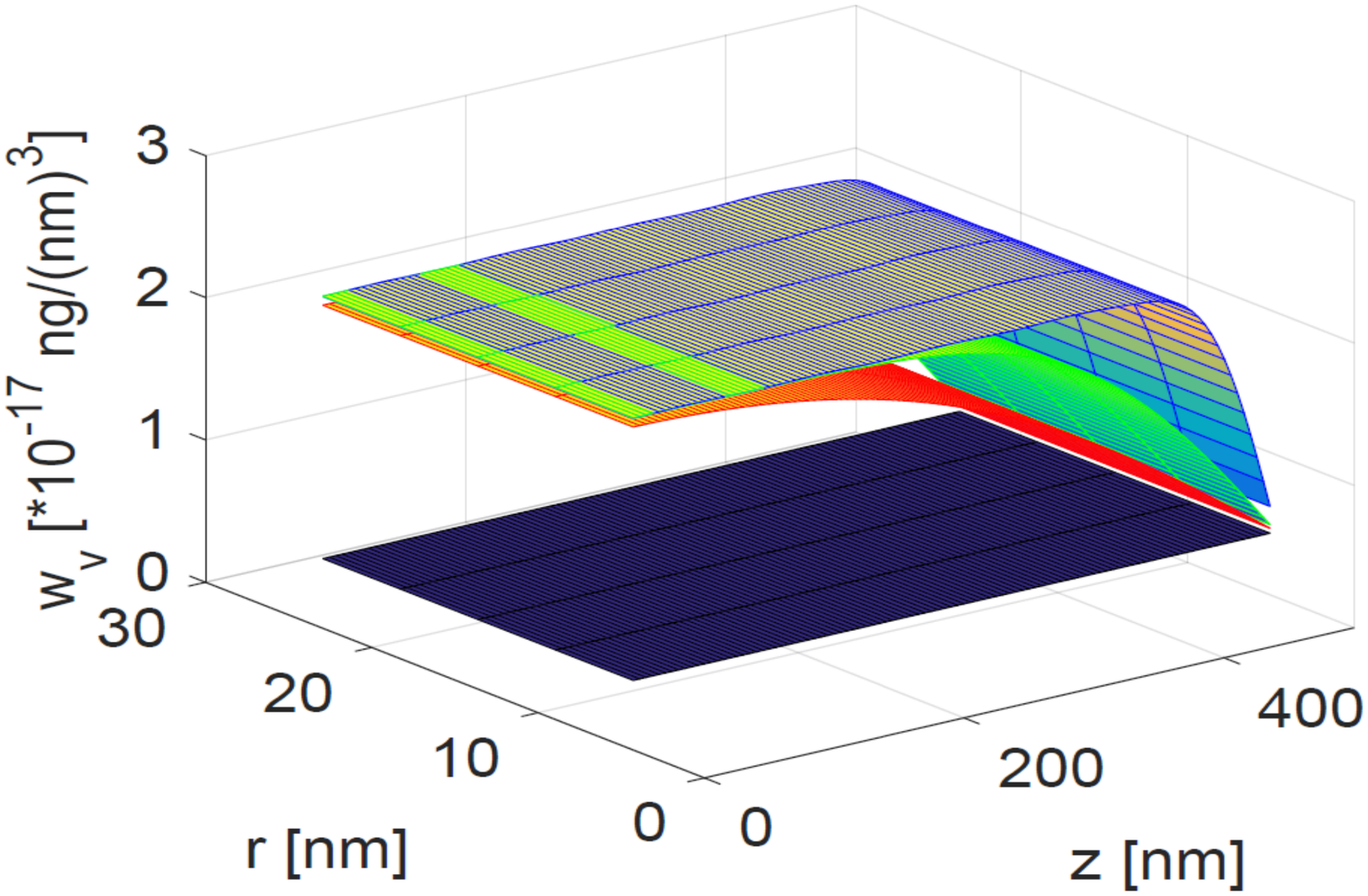}
\end{center}
\caption{Dynamics of the water vapor concentration for different time moments $t=6,60, 180$ and 65300 psec from up to down.}\label{Fig3}
\end{figure}

According to formula (\ref{eq06}), we drew the solution for the macro-model  (\ref{eq01})--(\ref{eq05}) that is depicted in the Fig. \ref{Fig3}. Fig. \ref{Fig3} consists of four surfaces that correspond to the water vapor concentration $w_v$ for four time moments $t=6, 60, 180$ and $65300$ psec.
   
\begin{figure}[H]
\begin{center}
\includegraphics[angle=-90,width=0.48\linewidth]{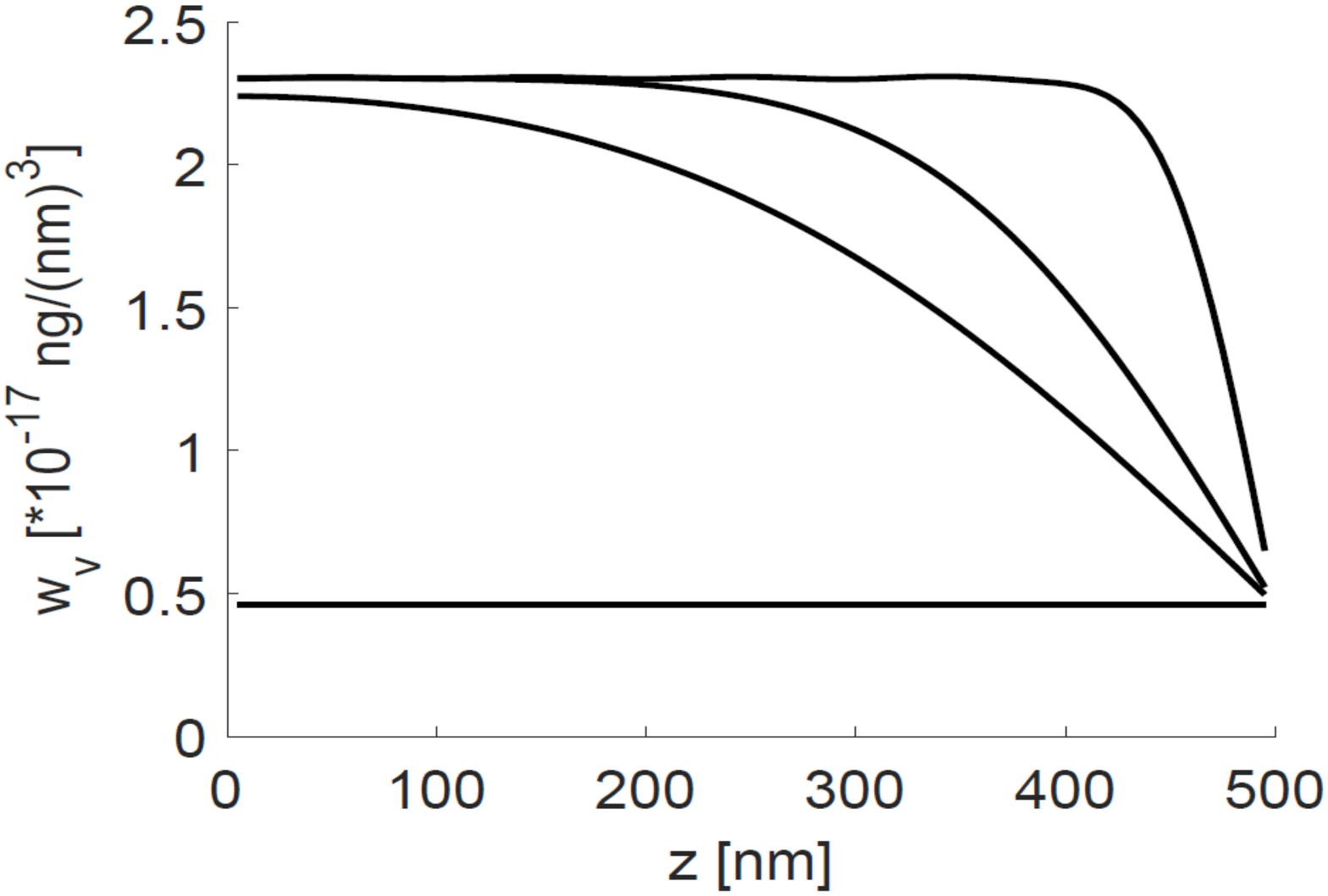}~
\includegraphics[angle=-90,width=0.48\linewidth]{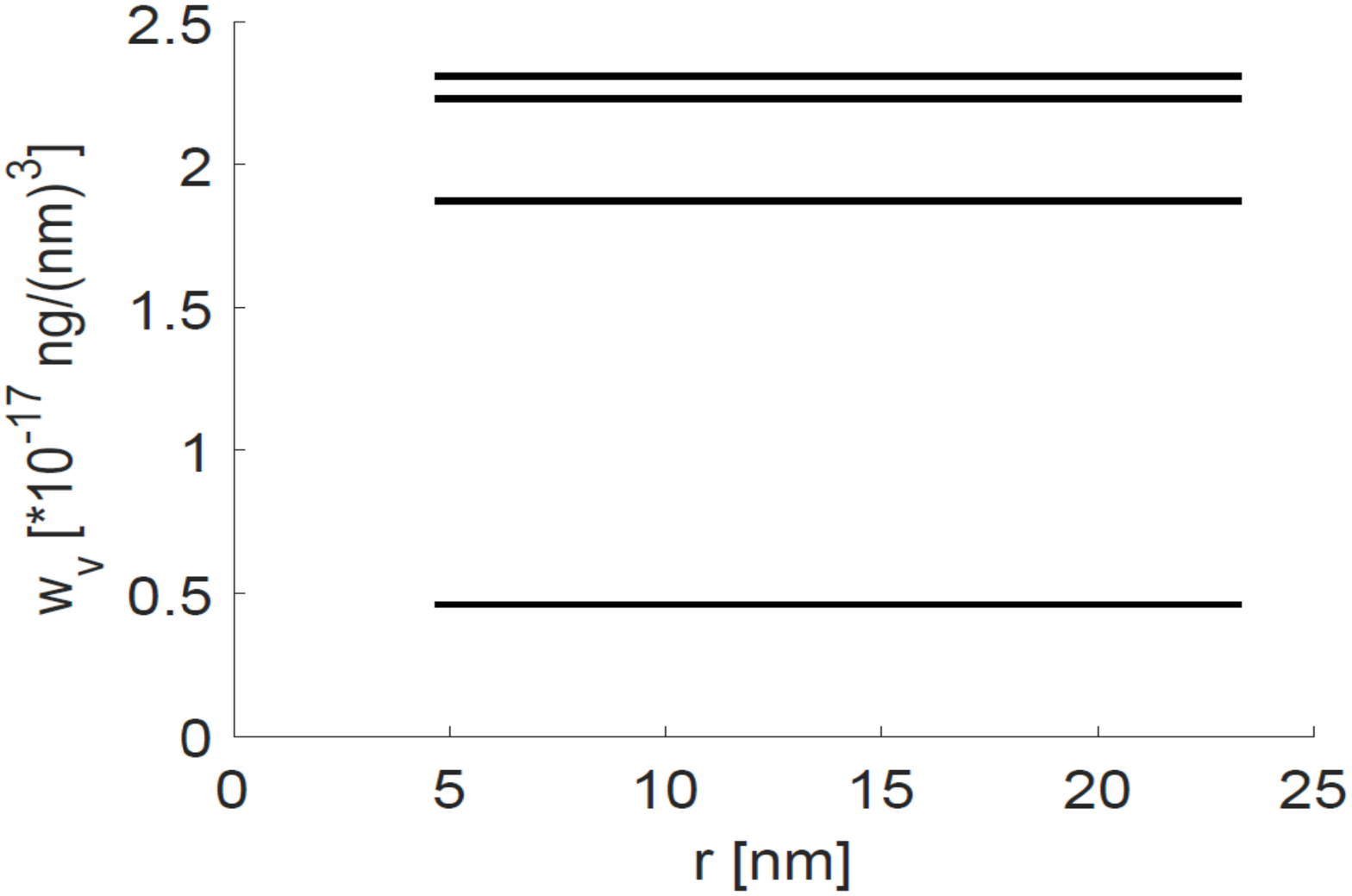}
\end{center}
\caption{Dynamics of the water vapor concentration for different time moments $t=6, 60, 180$ and 65300 psec from up to down at cross-section $r=r_0/2$ (left) and at cross-section $z=z_0/2$ (right).}\label{Fig4}
\end{figure}

Next, the Fig. \ref{Fig4} shows two cross-sections of the Fig. \ref{Fig3}. Fig. \ref{Fig4} (left) relates to the cross-section $r=r_0/2=14.25$ nm and Fig. \ref{Fig4} (right) relates to the cross-section  $z=z_0/2=250$ nm. Both figures (left and right) show that the water vapor concentration is decreasing in the cylindrical pore i.e. the pore is drying.  

\section{Comparison of microscopic and macroscopic models}\label{comparison}
\begin{figure}[H]
\center{
\includegraphics[angle=-90,width=0.48\linewidth]{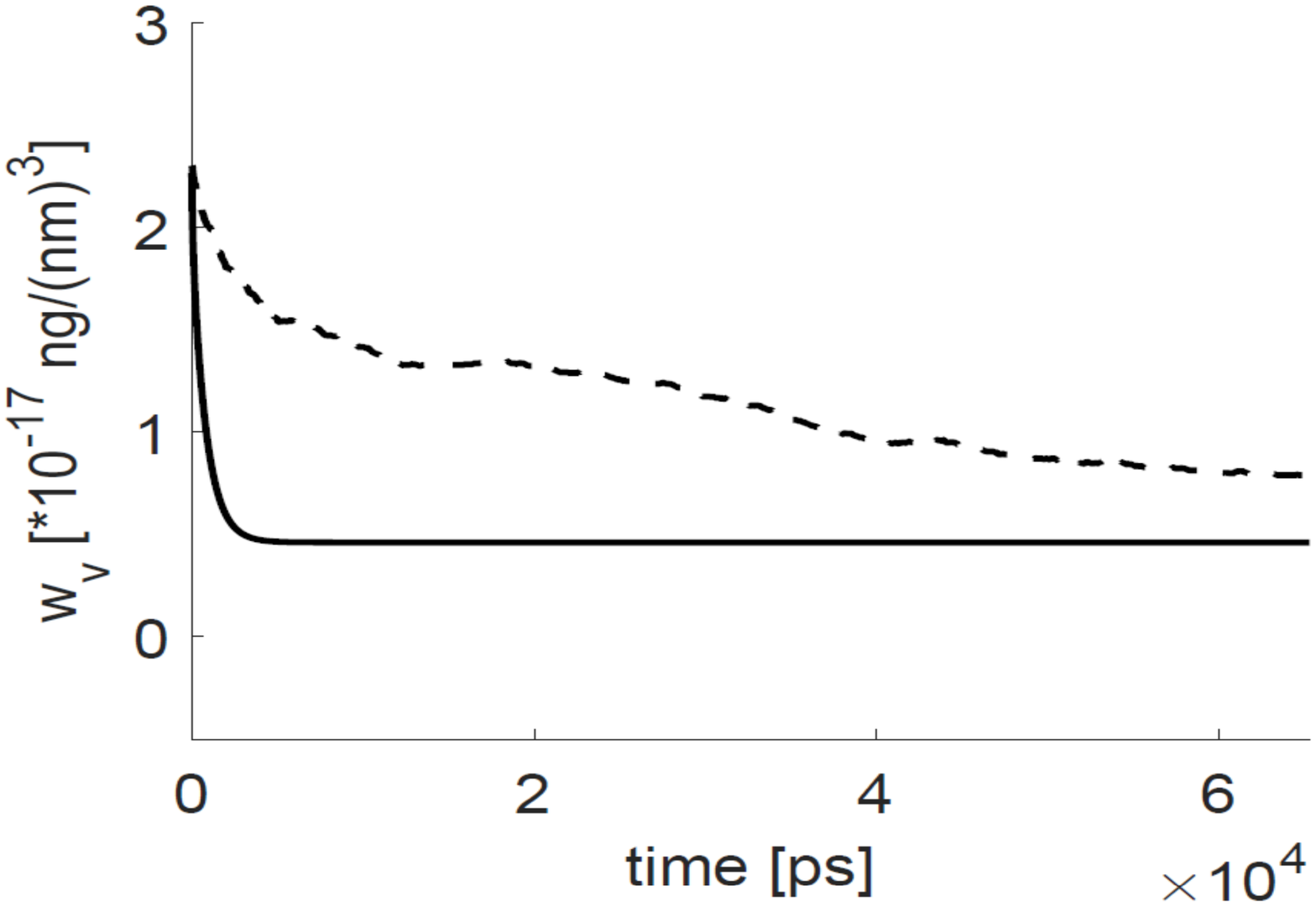}
}
\caption{The dynamics of water vapor concentration (WVC) obtained by diffusion equation (lower curve) versus WVC obtained by molecular dynamics (upper curve).}
\label{wv-rho}
\end{figure}

Finally, we compare the space mean value 
$$
\frac{1}{\pi r_0^2 z_0}\int_{0}^{r_0}\int_{0}^{2\pi}\int_{0}^{z_0}w_v(r\cos\varphi,r\sin\varphi,z,t)r drd\varphi dz
$$
of water vapor concentration (\ref{eq06}) for macro model with the density obtained by micro model. The results are in the Fig.\ref{wv-rho}.   

\section{Conclusions}
Our investigations allow to affirm that an approach based on combination of diffusion coefficients determination by means of molecular dynamics and further application of  these coefficients in macro model computations is useful for accuracy increasing of water-pore interaction description if the most accurate diffusion coefficient map is used. In this case the simulation of moisture transfer through porous media by widely used in material science software such as WUFI, COMSOL etc. has to  be more accurate without loss of calculation efficiency.

\end{document}